\RequirePackage{fix-cm}
\documentclass[smallextended]{svjour3}       % onecolumn (second format)
\smartqed  % flush right qed marks, e.g. at end of proof
\usepackage{graphicx}
\usepackage{amssymb}
\usepackage{float}
\usepackage{subfig}
\newcommand{\aap}{A\&A, }

\newcommand{\planss}{Planet. Space Sci., }
\newcommand{\solphys}{Sol.~Phys., }
\begin{document}
\title{A systematic analysis of the XMM-Newton background: III. Impact of the magnetospheric environment.}
%\subtitle{Do you have a subtitle?\\ If so, write it here}
\titlerunning{Impact of the magnetospheric environment on the XMM-Newton background}  
%\titlerunning{Short form of title}        % if too long for running head

\author{Simona Ghizzardi$^{1}$ \and
        Martino Marelli$^{1}$ \and
        David Salvetti$^{1}$ \and
        Fabio Gastaldello$^{1}$ \and
        Silvano Molendi$^{1}$ \and
        Andrea De Luca$^{1,3}$ \and
        Alberto Moretti$^{2}$  \and
        Mariachiara Rossetti$^{1}$ \and  
        Andrea Tiengo$^{1,3}$
}

\authorrunning{Simona Ghizzardi et al} % if too long for running head

\institute{S. Ghizzardi
               \at $^{1}$ INAF-IASF Milano, via E. Bassini 15, I-20133 Milano, Italy\\
           \email{simona@iasf-milano.inaf.it}         
           \and
%           \at 
$^{2}$ INAF-Osservatorio Astronomico di Brera, via Brera 28, I-20121 Milano, Italy
           \and\\
%           \at 
$^{3}$ Istituto Universitario di Studi Superiori, piazza della Vittoria 15, I-27100 Pavia, Italy}
             
\date{Received: date / Accepted: date}
% The correct dates will be entered by the editor

\maketitle

\begin{abstract}
A detailed characterization of the particle induced background is fundamental for many of the
scientific objectives of the Athena X-ray telescope, thus an adequate knowledge of the background that 
will be encountered by Athena is desirable.
Current X-ray telescopes have shown that the intensity of the particle induced 
background can be highly variable.
Different regions of the magnetosphere can have very different environmental conditions,
which can, in principle, differently affect the particle induced background  
detected by the instruments.
We present results concerning the influence of the magnetospheric environment on the
background detected by EPIC instrument onboard XMM-Newton through the estimate
of the variation of the in-Field-of-View background excess along the XMM-Newton orbit.
An important contribution to the XMM background, which may affect the Athena background as well, comes from soft proton flares. 
Along with the flaring component a low-intensity component is also present. We find that both show modest variations
in the different magnetozones and that the soft proton component shows a strong trend with the distance from Earth.


% \PACS{PACS code1 \and PACS code2 \and more}
% \subclass{MSC code1 \and MSC code2 \and more}
\end{abstract}

\section{Introduction}
\label{intro}

The characterization of the background in X-ray observations is a major concern for
astronomers interested in studying faint and diffuse sources. The European Photon Imaging Camera 
(EPIC) on board XMM-Newton does not provide an exception to this situation. 
Its instrumental background can be divided into electronic noise and particle-induced background 
(see \cite{Gasta_ahead:2017} for a detailed description). The latter component has two main contributions: 
an ``unfocused'' component, caused by high-energy particles ($E >$ MeV) which are able to reach 
also the unexposed regions of the field-of-view (FOV) and and a ``focused'' component, 
which causes an excess of signal only in the part of the FOV exposed to the sky and is 
usually associated to the so called ``soft protons''. These low-energy particles (a few tens of keV) 
are somewhat focused by the telescope and do not produce signal in the unexposed corners of the FOV. 
When the satellite encounters in its orbit a cloud of such particles a sudden and highly variable 
count-rate excess is detected (``soft proton flares''), which hampers the scientific exploitation 
of the data. These particles are likely accelerated in the Earth magnetosphere and therefore the 
intensity of the particle induced background may depend on the magnetospheric environment during 
the observations. 
The orbit of XMM-Newton is highly elliptical (with an apogee of about 115,000 km and a perigee
of about 6,000 km from Earth) and crosses regions of 
the magnetosphere with different properties in terms of strength and orientation of the magnetic 
field, speed and density of the particles etc. It goes from the radiation belts near the perigee, 
through the magnetoplasma and magnetotail, to the magnetosheath and eventually out of the bow shock 
into the solar wind.
% The XMM-Newton orbit is a highly eccentric, elliptical orbit . During each orbit, the satellite transits through the Earth
% magnetosphere, the region around the Earth influenced by its magnetic field. Different regions of the
% magnetosphere can have very different environmental conditions, depending on the strength and the
% orientation of the magnetic field, the speed and the density of the particles etc. The different conditions
% encountered by XMM-Newton can, in principle, differently affect the particle background detected by
% EPIC. While traveling along its orbit, XMM-Newton crosses these different
% magnetospheric environments, from the radiation belts near the perigee, through the magnetoplasma
% and magnetotail, to the magnetosheath and eventually out of the bow shock into the solar wind.
XMM-Newton data are therefore very useful to test the dependence of the induced particle background
in different magnetospheric environments.

This work is part of a wider project that aims to characterize the effects
of focused and unfocused particles on X-ray detectors through the analysis of XMM-Newton data. 
In this paper we focus on the impact of the magnetospheric enviroment on the XMM-Newton 
background components that cause an excess count rate in-the-field-of-view ({\it inFOV}) 
with respect to the unexposed corners (focused background component; {\it outFOV}). Hence to investigate 
and quantify the background we make use of
the difference {\it inFOV}-{\it outFOV} rate (see Sec \ref{sub:ambient}). 
Complementary results are presented in companion papers: details about the data reduction, cleaning and filtering are 
provided in \cite{Marelli_ahead:2017}; in \cite{Salvetti_ahead:2017} we provide a characterization of the 
focused background component. Finally in \cite{Gasta_ahead:2017} we present results about the origin of the unfocused 
particle background and about the focused soft protons background.
This work has been developed within the AREMBES\footnote{http://space-env.esa.int/index.php/news-reader/items/AREMBES.html}
(Athena Radiation Environment Models and X-ray Background Effects Simulators) project, 
a ESA R\&D Activity. The data sample construction and reduction has been performed thanks to the synergy with 
EXTraS\footnote{http://www.extras-fp7.eu} (Exploring the
X-ray Transient and variable Sky; \cite{De_Luca.ea:15}), a EU-FP7 project.

\section{The dataset}
\label{sec:dataset}
We adopt the largest XMM-Newton data set ever analysed based on the entire XMM-Newton archive. It collects $\sim 100$Msec of data 
from observations performed between 2000-2012 (revolution 35 to 2330). 
The description of the sample and the reduction and cleaning procedures are provided in \cite{Marelli_ahead:2017}.
In addition, we reject periods that are classified as ``SEP contaminated'' to avoid eventual unwanted biases;
the list of all the SEP contaminated periods is provided by the ESA Solar
Energetic Particle Environment Modelling (SEPEM) application server{\footnote{http://dev.sepem.oma.be/help/event\_ref.html}}.
After the removal of the time intervals affected by SEP events, the sample reduces to 87.8 Msec of
cleaned data.

% SEP events are related to solar flares and mass coronal ejection and
% occasionally they can induce a notable increase of the EPIC background level.
%The influence of SEP
%on the EPIC background will be extensively addressed in {\bf Gasta} and the induced
%contamination is not systematic and obvious. 
% We use the ESA Solar
% Energetic Particle Environment Modelling (SEPEM) application server 8 to obtain the
% list of all SEP contaminated periods.

\section{Method}
\label{sec:method}
\subsection{Partition of the magnetosphere into magneto-zones}
\label{sub:m-zones}

\begin{figure*}
   \includegraphics[width=0.45\textwidth]{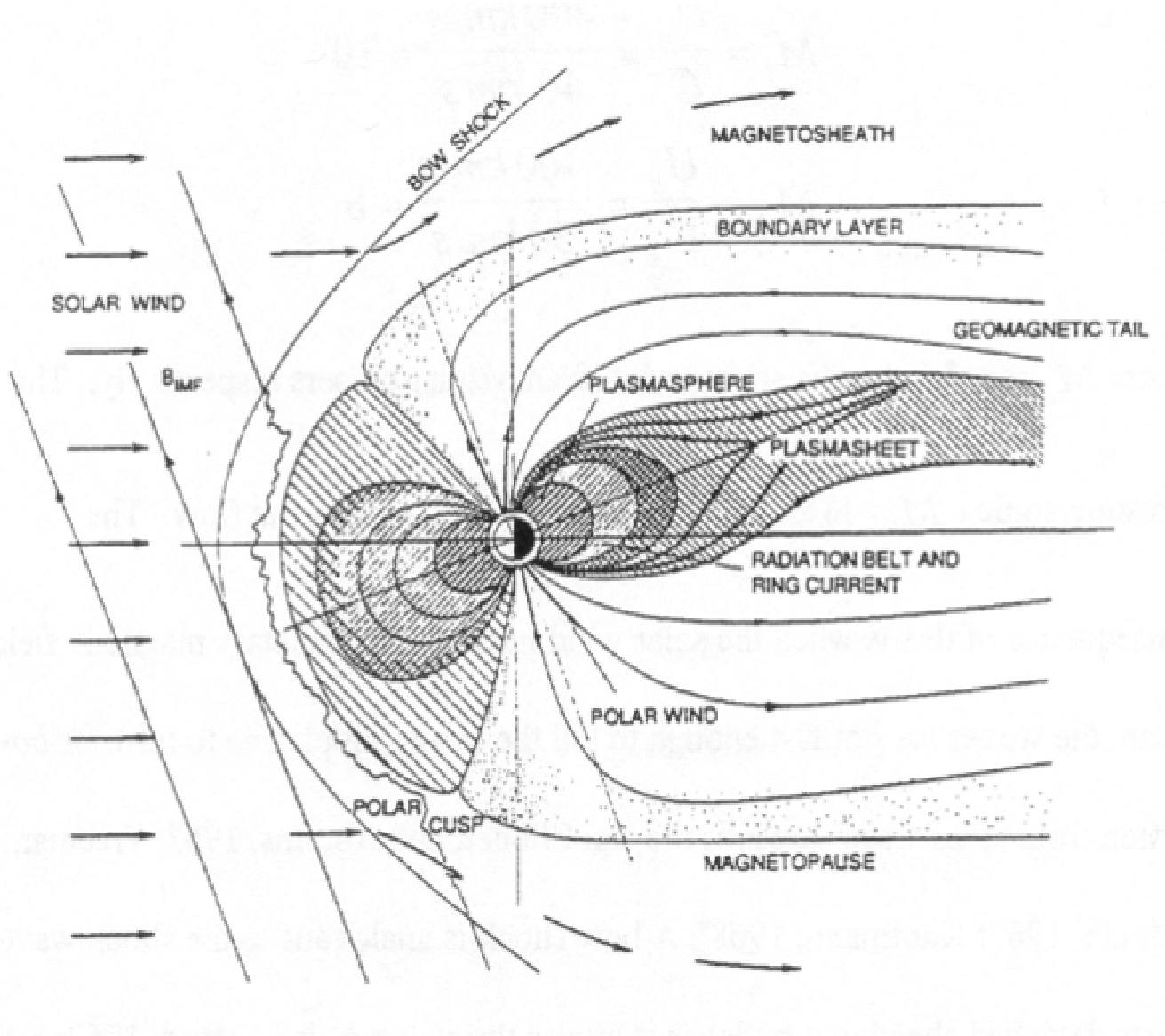}
   \includegraphics[width=0.45\textwidth]{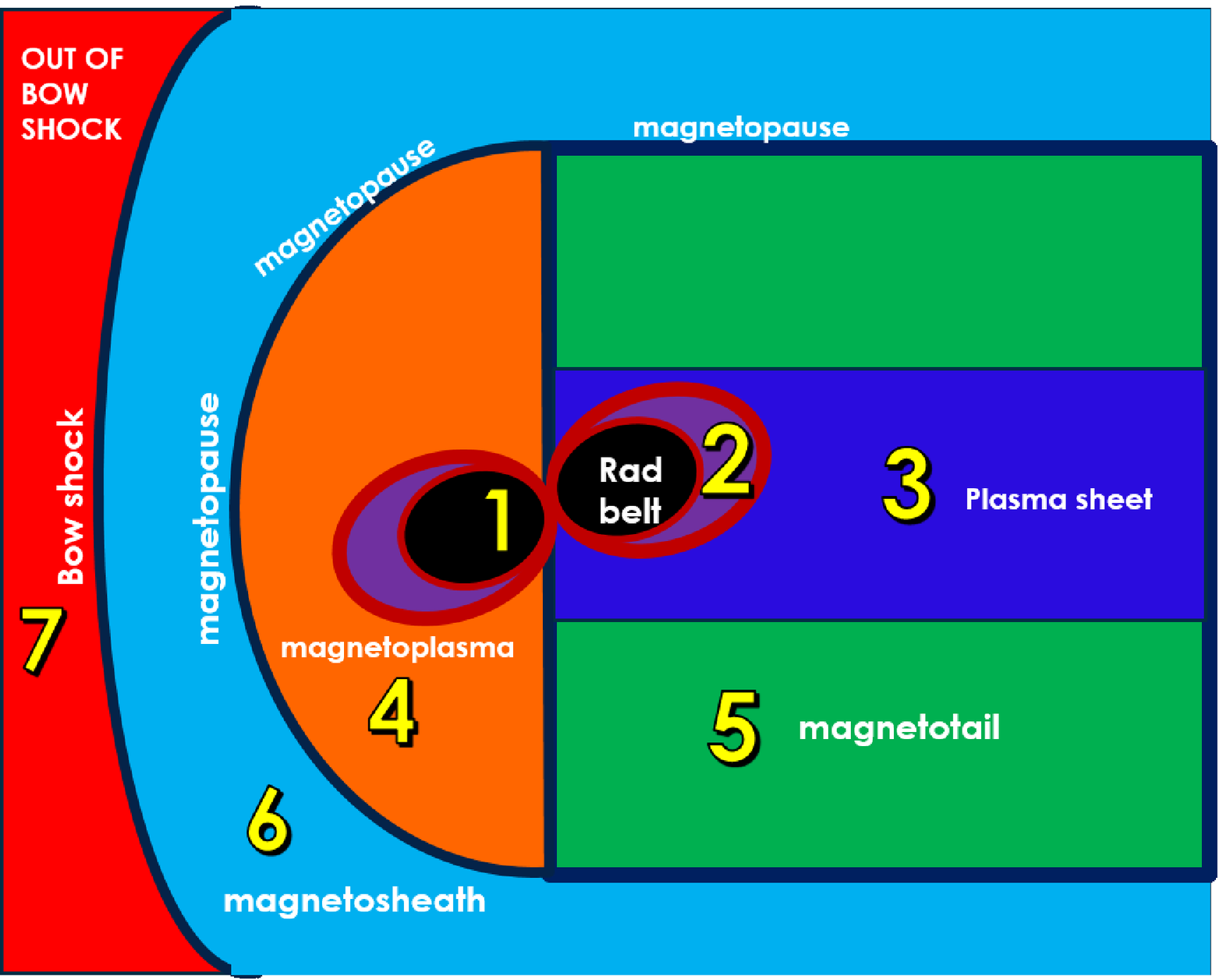}
% figure caption is below the figure
\caption{A schematic view of the magnetosphere of the Earth (left panel) and our simplified division of the
magnetosphere into 7 magneto-zones (right panel). The color code represented here will be adopted throughout the
paper.}
\label{fig:zones}       % Give a unique label
\end{figure*}

The terrestrial magnetosphere prevents most of the solar wind from hitting the Earth, although some
energetic particles can enter it. In Figure \ref{fig:zones} (left panel), we provide a schematic representation of the
Earth magnetosphere. 

The outermost layer of the magnetosphere is the bow shock; it forms when the
supersonic solar wind encounters the Earth magnetic field. The solar wind across the bow shock surface
is then heated up and slowed down by the Earth's magnetic field which acts like an obstacle. As a
consequence, the solar wind starts flowing around the obstacle forming the magnetopause, a surface
which divides the terrestrial magnetic field from the solar wind that flows around it. 

We adopt a simplified characterization of the Earth magnetosphere (see Figure \ref{fig:zones}, right panel) and
divided the magnetosphere into 7 typical magneto-zones. 
The Van Allen radiation belts are modeled through the L-shell model by \cite{McIlwain1961} $R = L \cos{2\lambda}$, 
where: $R$ is the radial coordinate of the field line in units of Earth radii $(R_E = 6371 {\rm km})$; $\lambda$ is the magnetic
latitude and the L-shell parameter is  $L= {R_0/R_E}$; $R_0$ is the intersection of the field line with the geomagnetic Equator. 
Variables are defined in the geocentric solar magnetospheric system (GSM). Since the external boundary of the radiation belts
is highly variable, we split this region into two different zones: 
the ``radiation belts'' (\#1) inside the $L = 4$ shell and the ``radiation belts exit'' (\#2):
the region between $L=4$ and $L=6$ where the satellite exits the belts and enters the outer
magnetospheric ambient.
The plasma sheet (\#3) is assumed to be a cylindric region centered on the Earth-Sun line,
with the axis parallel to the ecliptic plane in the anti-Sunward direction, with radius $R = 5 R_E$ \cite{Rosenq2002}.

Inside the magnetopause, the magnetic field lines have a different shape on the nightside and dayside
regions. In the dayside region, magnetic field lines are closed, distorted and
compressed by the pressure of the solar wind. Conversely, in the nightside regions the magnetic field
lines are stretched and open.
We divide the area inside the magnetopause into two different sectors: the anti-
Sunward region is known as magnetotail (\#5) and we dub ``magnetoplasma'' the Sunward zone (\#4).
To model these regions, we use a simple model \cite{KS2008} for the magnetopause
radius in the dayside direction
\begin{equation}
 R_{MP} =
{14.21 \over {1 + 0.42\cos{\theta}}}
\end{equation}

where the distance $R_{MP}$ is in $R_E$ units and $\theta$ is the angle from the Earth-Sun line. Coordinates are in the
geocentric solar ecliptic (GSE) system. 
On the nightside, we assume the magnetopause to be a cylindrical surface, with radius $R = 14.21 R_E$,
with the cylinder axis parallel to the ecliptic plane and centered on the Earth-Sun line.

The magnetosheath (labelled as \#6) is the plasma region between the bow shock and the magnetopause in which the
shocked solar wind is heated and slowed down from supersonic to subsonic speeds. 
The boundaries for this 
magneto-zone  are the magnetopause surface and the bow
shock surface that we model following \cite{KS2008}:
\begin{equation}
 R_{BS} = {22.74 \over {1 + 0.75 \cos{\theta}}}
\end{equation}

where the distance $R_{BS}$ is in $R_E$ units and $\theta$ is the angle from the Earth-Sun line. 

We finally label as magneto-zone \#7, the regions out of the bow shock when the satellite is outside the
magnetosphere and embedded in the solar wind. 

The description used in our analysis for the
magnetosphere is clearly simplified and the model neglects possible time variations of the shape and 
boundaries of the magneto-zones: solar wind speed and pressure vary with time and eventually compress the
magnetopause and bow shock surfaces changing their boundaries. It should also be noted 
that region boundaries are not sharp edges and these zones are
not strictly distinct, instead there may be smooth transitions from one region to another. 
However, also
thanks to the very large quantity of data available, this simple description of the magnetosphere is
appropriate to study in a statistical way how the various magnetospheric conditions can affect the XMM
particle background.

\subsection{XMM-Newton orbit segmentation}
\label{sub:segment}
\begin{figure*}
 \includegraphics[width=0.5\textwidth]{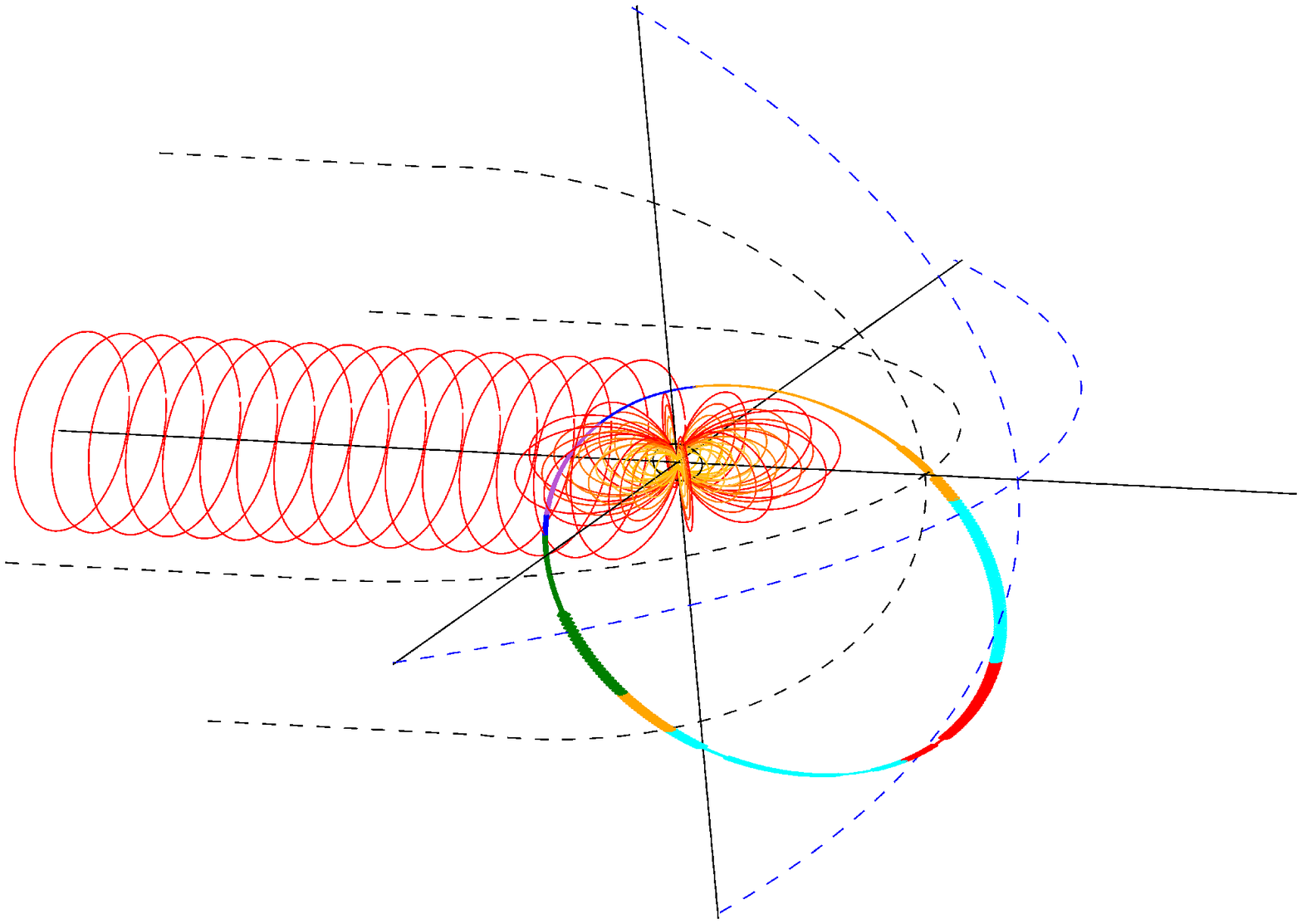}
 \includegraphics[width=0.5\textwidth]{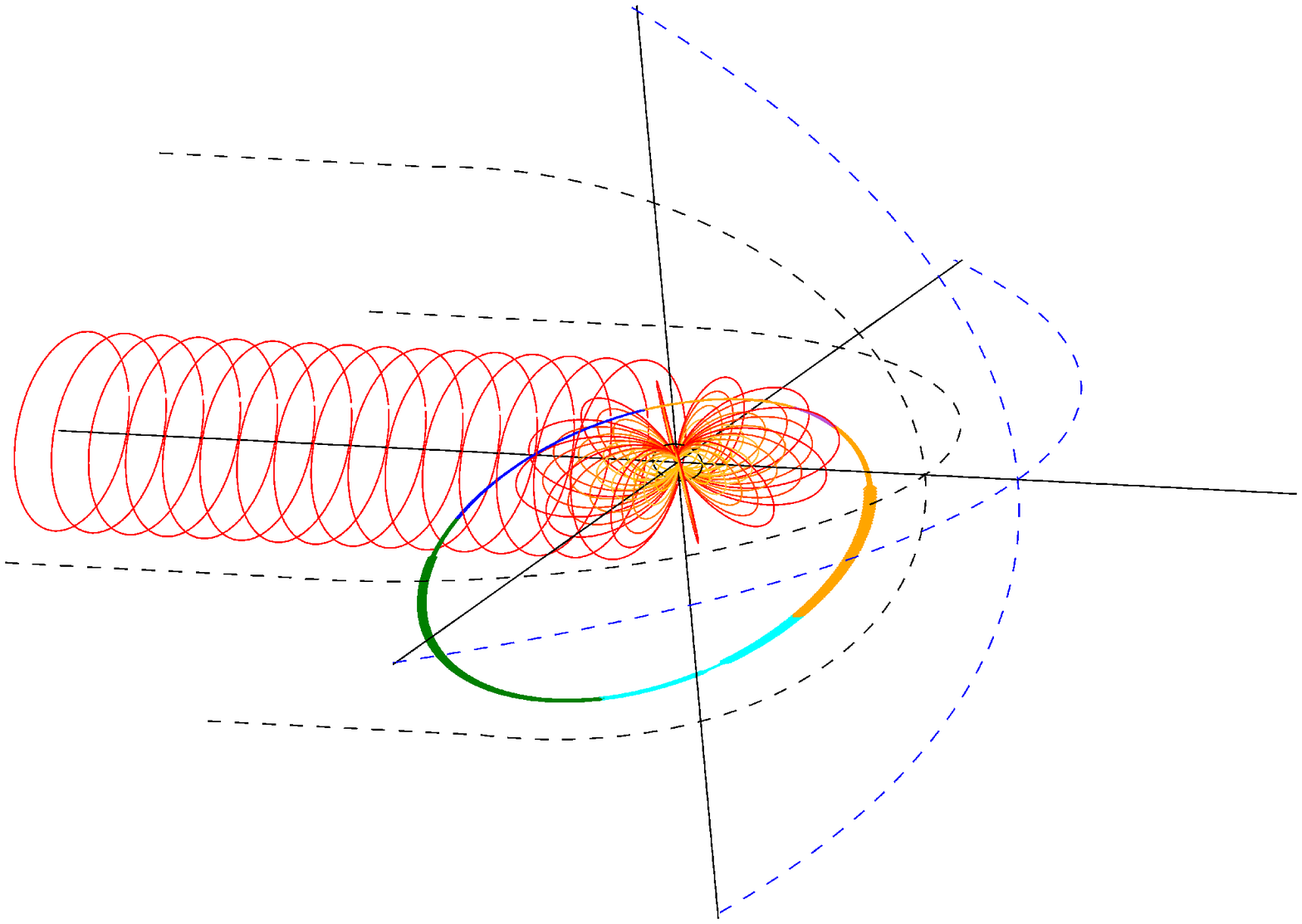}
\caption{XMM-Newton orbit for revolution 1016 (left panel) and 1466 (right panel). The coordinate grid is in Geocentric
Solar Ecliptic (GSE) coordinate system, with the Earth in the origin and the Sun located at the end of the X-axis at the
right-side of the plot; XY plane is the ecliptic plane. Radiation belts (closed lines arount the
Earth) are plotted for $L$ in the range $L= 2$(yellow) to $L=6$(red) with colors with orange tones for intermediate values of L.
Red cicle mark the plasma sheet and black and blue dashed lines are the projection of the magnetopause and bow shock
surface respectively. The XMM-Newton orbit segments are plotted using the color codes defined in Figure \ref{fig:zones}.}
\label{fig:segment}
\end{figure*}

Our sample includes data from revolution 35 to 2330. For each revolution, we derive the XMM-Newton
orbit and divide it into segments according to the magnetosphere environment crossed while travelling.
Then, for each revolution and for each magneto-zone we find the Good Time Intervals (GTI) that can
be used to filter the sample data and analyze the background region by region.
In Figure \ref{fig:segment} we plot, as an example, a 3D representation of the XMM-Newton 
orbits during revolution 1016 (26-27 June 2005) and revolution 1466 (10-11 December 2007). 
Closed lines
around the Earth track the torus of the radiation belts whose orientation varies in time 
due to seasonal and daily motion of the Earth's dipole tilt
angle. To derive magnetic axis inclination changes, we use the SolarSoftware (SSW) IDL package \cite{FH98}, 
where the dipole axis position is calculated according to the International
Geomagnetic Reference Frame (IGRF) model, as described in \cite{FH2002}.
The plasma sheet cylinder is represented through a series of red circles, extending in the
anti-Sun side; black and blue dashed lines reproduce the magnetopause and the bow shock surface
respectively.

The orbit segments are plotted using the color code defined in Figure \ref{fig:zones} and the orbit parts where
EXTraS data are available are plotted with a thick line. EXTraS data generally cover only a fraction
of the orbit. The lack of data
during the revolution can be due to various reasons. First of all, EPIC cameras are closed at low
altitudes to avoid damage from exposure to soft protons during the passages through the radiation
belts: XMM has a minimum observation altitude of 40,000 km. This is responsible of missing data at the
beginning and at the end of each orbit. Observations can be missing for corrupted or bad data or could
have been rejected from the EXTraS archive \cite{Salvetti_ahead:2017}. In addition, gaps are present during
slew transitions from an observation target to another. 
The portions of the orbits where EXTraS data are not available have been reconstructed 
using the information available in the Radiation Monitor page of the XMM-Newton 
website\footnote{http://www.cosmos.esa.int/web/xmm-newton/list-of-tc-radmon} where
fits files containing orbit parameters can be retrieved. In these files, the XMM-Newton orbit status is stored 
with a 1 second cadence and their processing can be very time-consuming. When available, 
we used Trend Data in HEASARC\footnote{https://heasarc.gsfc.nasa.gov/docs/xmm/xmmhp\_trend.html}, as
they provide XMM orbit parameters with a 64 sec cadence. We thus use Radiation Monitor orbit files only
when Trend Data are missing.

XMM-Newton spends most of the time
south of the ecliptic plane. The direction of the orbit and the
apogee position change during the year. Depending on the season, the orbit extends toward the Sun,
with the apogee eventually exiting the bow shock surface (like in the left panel of Figure \ref{fig:segment}) 
or in the anti-Sun direction, keeping completely inside the magnetotail and the magnetosheath (right panel of
Figure \ref{fig:segment}).

\begin{figure*}
 \includegraphics[width=0.5\textwidth]{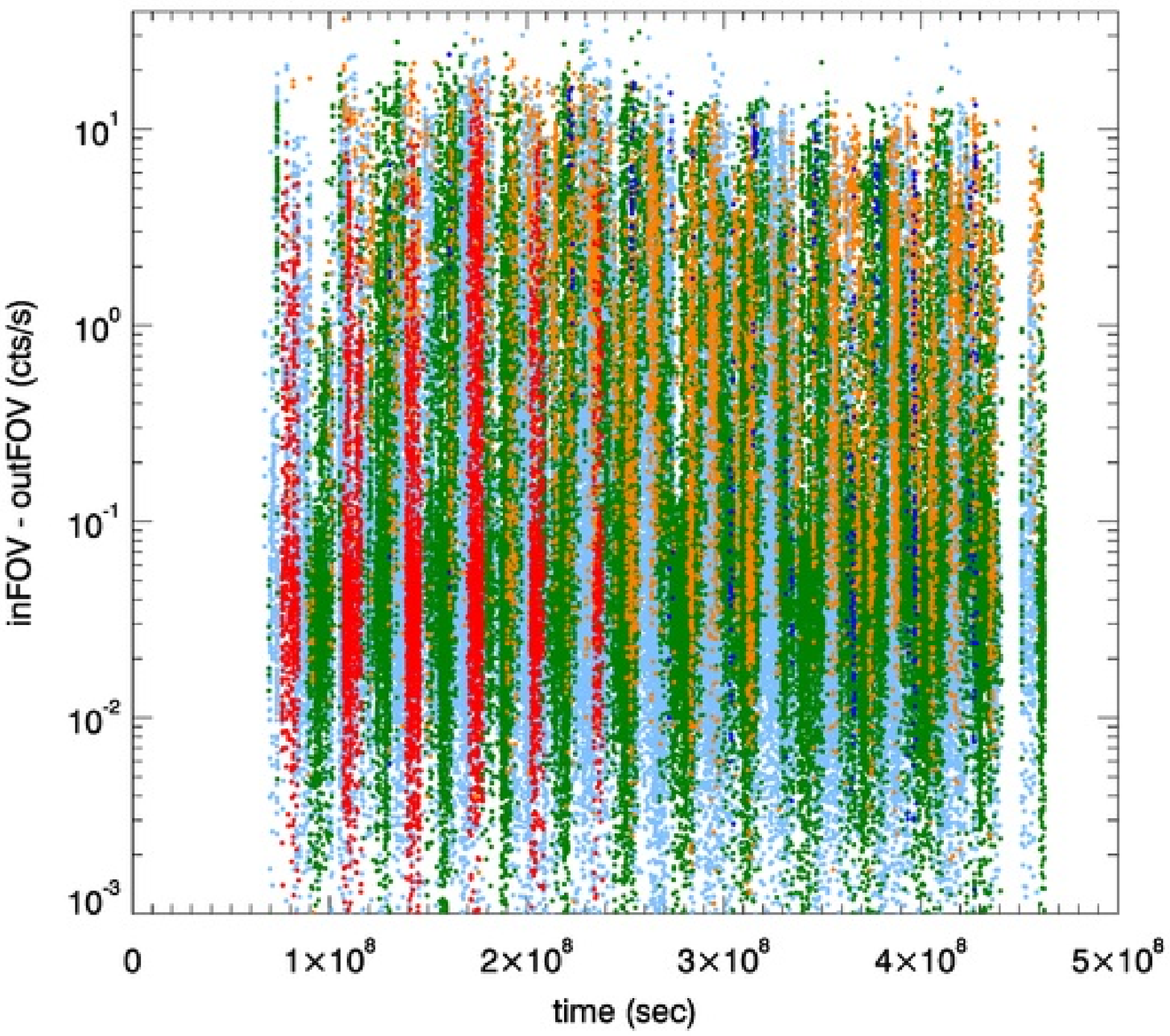}
  \includegraphics[width=0.5\textwidth]{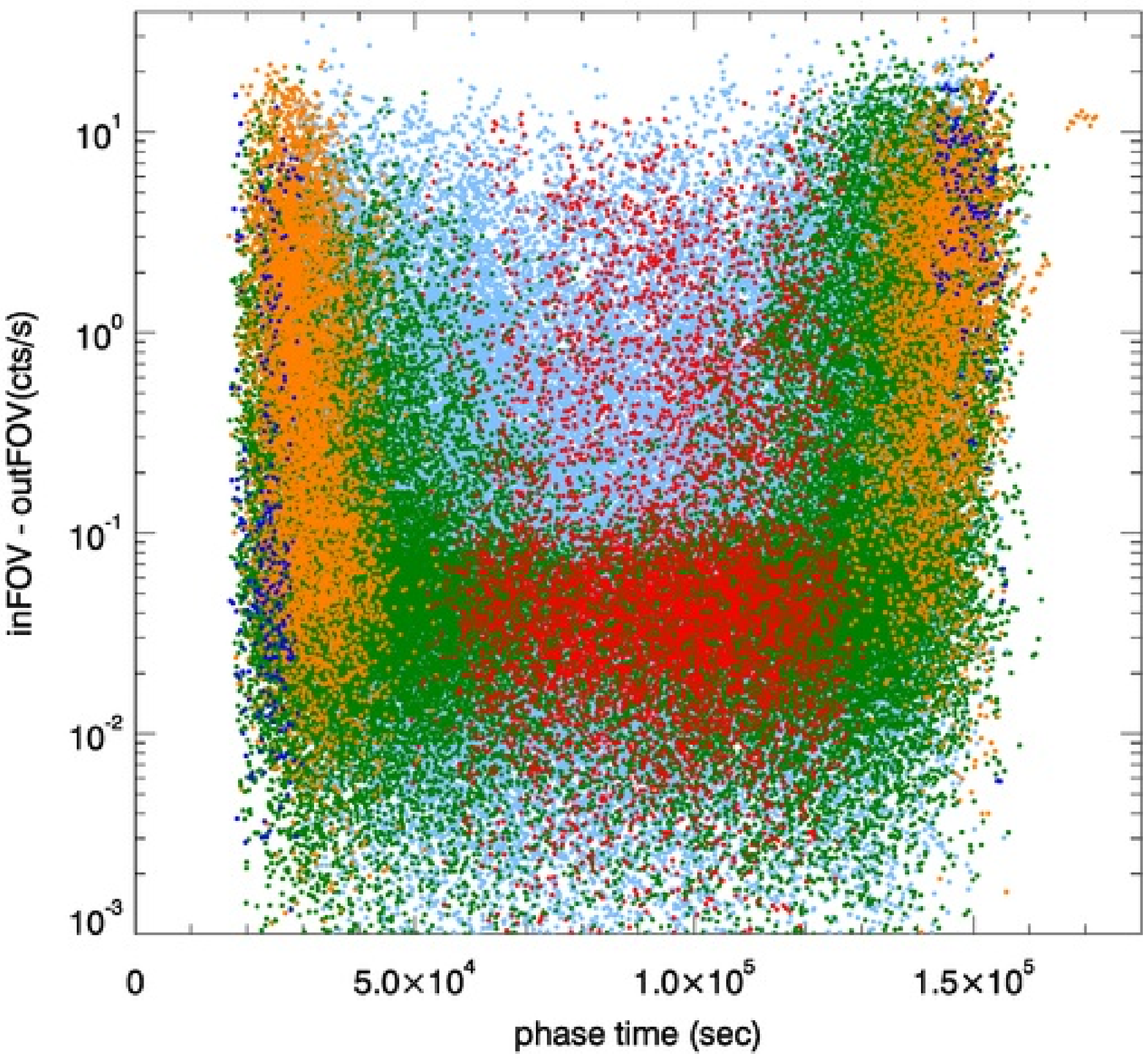}
 \caption{Background intensity ({\it inFOV}-{\it outFOV} rate, see Sec. \ref{sub:ambient}) 
 in the whole EXTraS sample versus time (left panel) and orbit phase (right).
 For each orbit, the phase time is the
time measured starting from the beginning of the orbit. The orbit is assumed to start at perigee. 
Colors mark the
different magneto-zones following the color codes of Figure \ref{fig:zones}}
\label{fig:full_lc-phase}
\end{figure*}

In Figure \ref{fig:full_lc-phase} (left panel) we plot the full lightcurve of the whole EXTraS sample, with colors marking the different
magneto-zones. In the first observation years, the out-of-bow-shock region (in red) is periodically
reached during the summer periods. Successively, namely after July 2005, the satellite is no more able
to reach this region, due to a gradual circularization of the orbit and to variations of its inclination angle.

\begin{table}
 \caption{We report the total time (and the corresponding fractional value) spent by XMM-Newton in the different
magnetospheric zones. We show both the time scored by the Radiation Monitor (which roughly corresponds to the total
time effectively spent in each region) and the total time (with the corresponding fractional value) of EXTraS data available
in the same region.
}
\label{table:times}       % Give a unique label
% For LaTeX tables use
\centering
\begin{tabular}{l|c|c|c|c}
\hline\noalign{\smallskip}
Magneto-zone & \multicolumn{2}{c|}{Radiation Monitor} &  \multicolumn{2}{c}{EXTraS Archive} \\
\noalign{\smallskip}\hline\noalign{\smallskip}
& Time &  Fraction & Time &  Fraction  \\
& (Msec) & (\%) &  (Msec) &  (\%) \\
\noalign{\smallskip}\hline\noalign{\smallskip}
\#1 Radiation belts & 3.3 & 0.9 & 0.0 & 0.0 \\
\#2 Radiation belts exiting & 13.2 & 3.4 & 0.0 & 0.0 \\
\#3 Plasma sheet & 20.3 & 5.2 & 0.4 & 0.5 \\
\#4 Magnetoplasma & 58.4 & 15.0 & 5.1 & 5.8 \\
\#5 Magnetotail & 126.1 & 32.5 & 35.6 & 40.5  \\
\#6 Magnetosheath & 154.7 & 39.8 & 43.0 & 49.0 \\
\#7 Out of bow shock & 12.4 & 3.2 & 3.7 & 4.2 \\
\noalign{\smallskip}\hline
\end{tabular}
\end{table}

During the 13 years under analysis XMM recursively crosses all the magneto-zones. The fraction of
time spent in each ambient depends on the orbit geometry and inclination and on the extension of each
zone. In Table \ref{table:times} we report the time (and fraction) spent in each magneto-zone and the
corresponding amount of EXTraS data. Particularly interesting is the out-of-bow-shock region (\#7),
where the satellite is out of the Earth magnetosphere. 
{\ bf This region is of particular interest as it should be mostly free of background 
components produced within the
magnetosphere.}
XMM-Newton spent in the out-of-bow-shock zone only 3.2\% of the time with 3.7 Msec of EXTraS data
in this region. Most of the time is spent into the magnetosheath and the magnetotail. Little time is spent
into the plasma sheet. Because of its position (in the nightside and along the ecliptic plane) and its
thinness, the plasma sheet hosts the satellite only for about 5\% of the time with only 5 Msec of data
available. Due to required off time near the perigee, no data are available in regions \#1 and \#2. These
two magnetospheric regions will not be discussed further in this paper.

\section{Results}
\label{sec:results}

\subsection{XMM-Newton background rate and magnetospheric environment}
\label{sub:ambient}

As anticipated in Sec. \ref{intro}, we use the {\it inFOV}-{\it outFOV} rate to estimate the EPIC background, i.e. the difference between 
the count rate measured in the area where X-ray photons are focused ({\it inFOV}) and the count rate 
measured in the unexposed areas ({\it outFOV}) of the detector.
Starting from the lightcurve of the whole EXTraS sample, we derived the {\it inFOV}-{\it outFOV} rate 
versus the orbit phase (Figure \ref{fig:full_lc-phase}, right panel); each orbit lightcurve is plotted  
versus the time elapsed from the perigee position.
This provides a qualitative picture of the EPIC {\it inFOV}
excess background along the orbit. Many
events feature a high (say $\gtrsim 0.1$ cts/s) {\it inFOV}-{\it outFOV} rate which can occasionally rise up to $\sim$ 200
cts/s; these correspond to soft proton flares. However, the bulk of the data lies in the range $[0.01 - 0.1]$
cts/s where the low-intensity component of the background dominates.
A full comprehension of the origin of both the soft proton flares and of the low-intensity component is still lacking.
Soft proton flares can include components having different origins: solar energetic particles events (SEP) or
particles generated 
at the bow shock or inside the magnetosphere (e.g. in the radiation belts). 
A more detailed discussion about this issue is provided in 
\cite{Gasta_ahead:2017}.
The nature of the low-intensity component is still unclear: as discussed in 
\cite{Salvetti_ahead:2017}, it is probably not associated to soft protons
and may be due to Compton interaction of hard X-ray photons with the detector.

Since the perigee is the starting (and ending) point of the orbit, at the center of the plot
we find the events recorded at the apogee: here are concentrated the ``out of bow shock'' data
(in red). Apparently, the {\it inFOV}-{\it outFOV} rate here is slightly lower than in the other regions, with a lower
spread of data, although not free from soft protons flare events. 
Blue and orange dots, respectively labeling the plasma sheet and the magnetoplasma on the dayside, are 
located at the edges of the plot near the perigee at the beginning and at the end of the orbit. 
Indeed, the satellite lies in these areas just after exiting (or before
entering) the radiation belts. 
The {\it inFOV}-{\it outFOV} rate in these regions seems on average larger than elsewhere 
and the quantity of data in this region is low since the time spent in the plasmasheet and 
into the magnetoplasma is only 0.4 Msec and $5$ 
Msec. Hence, Figure \ref{fig:full_lc-phase} (right panel) provides two relevant results: 1) the presence of soft proton flares 
is not related to any particular magnetozone, and they are distributed throughout all the different regions, 
2) no portion of the orbit is free from soft proton flares.

To quantify the variation of the {\it inFOV} background excess in the different magnetospheric ambients 
we plot, in Figure \ref{fig:histo_fit}, the distributions of {\it inFOV}-{\it outFOV} rate for 
the five considered zones: distributions on the left column
are zoomed to low {\it inFOV}-{\it outFOV} values (0.1-0.3 counts/s)
for a better visualization of the low-intensity component, while the wider range is used in the panel on the right column to better inspect the tail extension.
The distributions show the presence of two main
contributions, confirming the qualitative picture provided by Figure \ref{fig:full_lc-phase}: 
1) the peaked Gaussian-like distribution at low count rates describes the low-intensity
component where the bulk of the data lies; 2) all the distributions feature a long tail toward high count
rate values, representing the flaring component. The wide extension of the tail is a symptom of the
importance of the flaring component, which, in all the magnetospheric regions, accounts for a notable
fraction of events: indeed the fraction of time when the background is affected by soft protons flares is
$\gtrsim$ $30\% - 40\%$ (see \cite{Salvetti_ahead:2017}).
Following \cite{Salvetti_ahead:2017} we fit the distributions using a Gaussian function in addition 
to a modified Lorentzian distribution $F(x)$ defined:

\begin{equation}
 F(x)= {{LN  x^{\Gamma_1}} \over{1+ \left|{{2(x-LC)} \over {LW}}\right|^{\Gamma_2}}} e^{-x/X_0}
\end{equation}

where $LN$, $LC$ and $LW$ are the normalization, the center and the width of the Lorentzian component; 
$X_0$ is the exponential cut-off and $\Gamma_1$, $\Gamma_2$ are
the two slopes.
The best fit functions for each magnetozone are overplotted in Figure \ref{fig:histo_fit}. We stress that the adopted model is purely 
phenomenological and there exists a strong correlation between the parameters. This requires that we consider uncertainties on model 
parameters with some caution.

The Gaussian peak derived from the fitting procedure is suitable to quantify
the low-intensity component contribution in the different magnetospheric ambients.

Best fit values for each magnetozone are reported in Table \ref{table:bestfit} (second column).

\begin{figure}
 \subfloat{\includegraphics[width=0.5\textwidth]{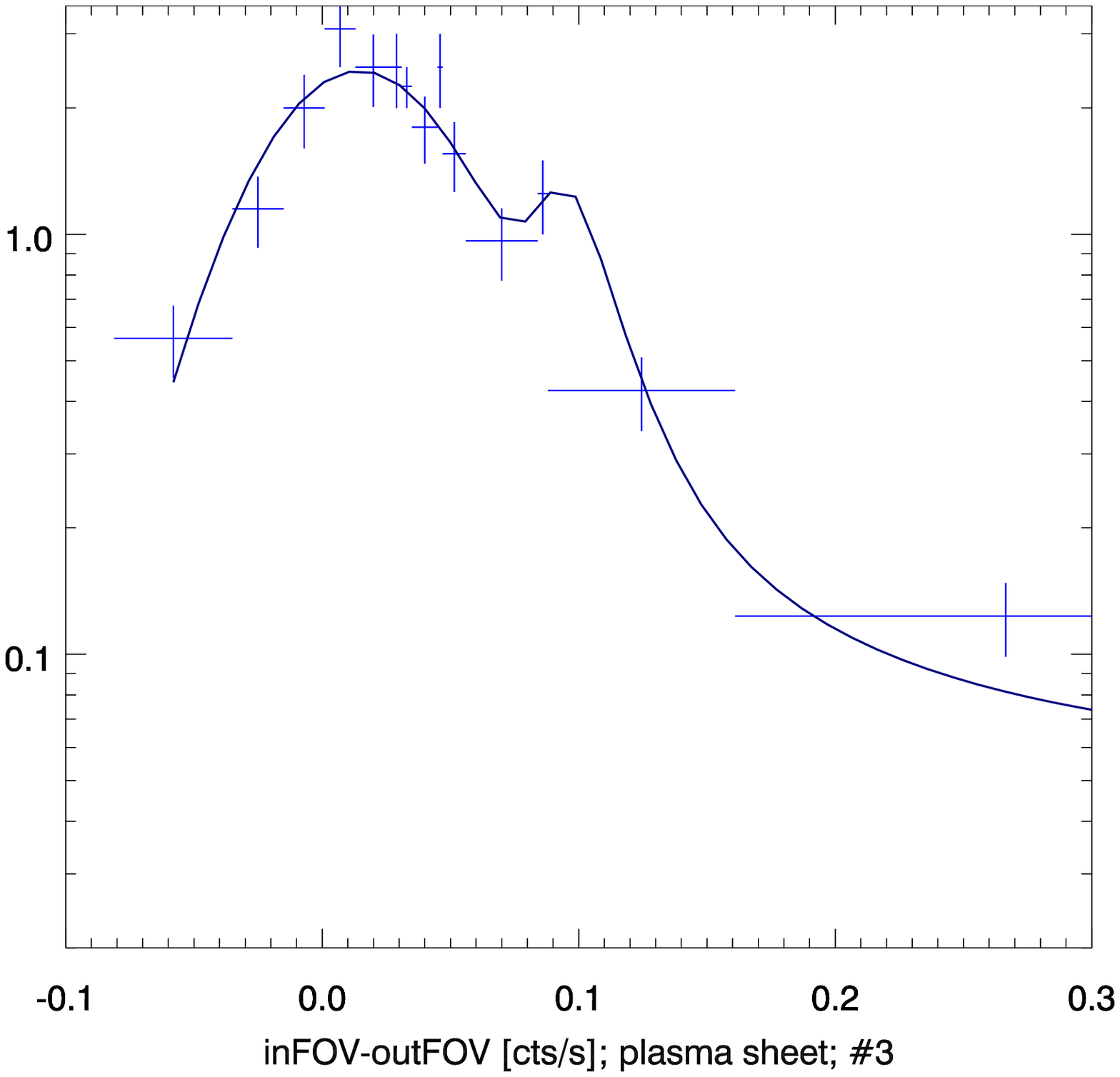} }
 \subfloat{\includegraphics[width=0.5\textwidth]{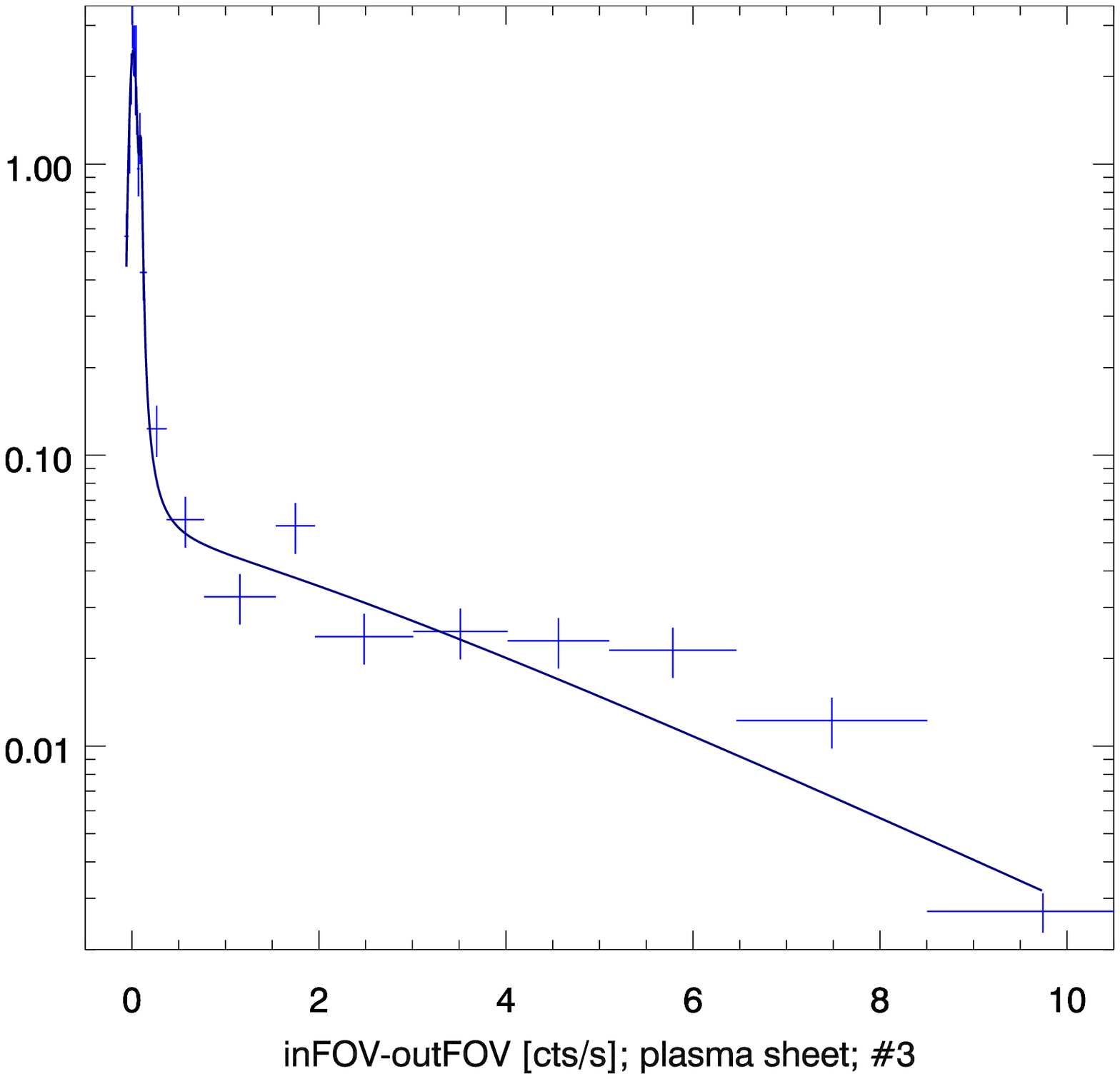} }\\
 \subfloat{\includegraphics[width=0.5\textwidth]{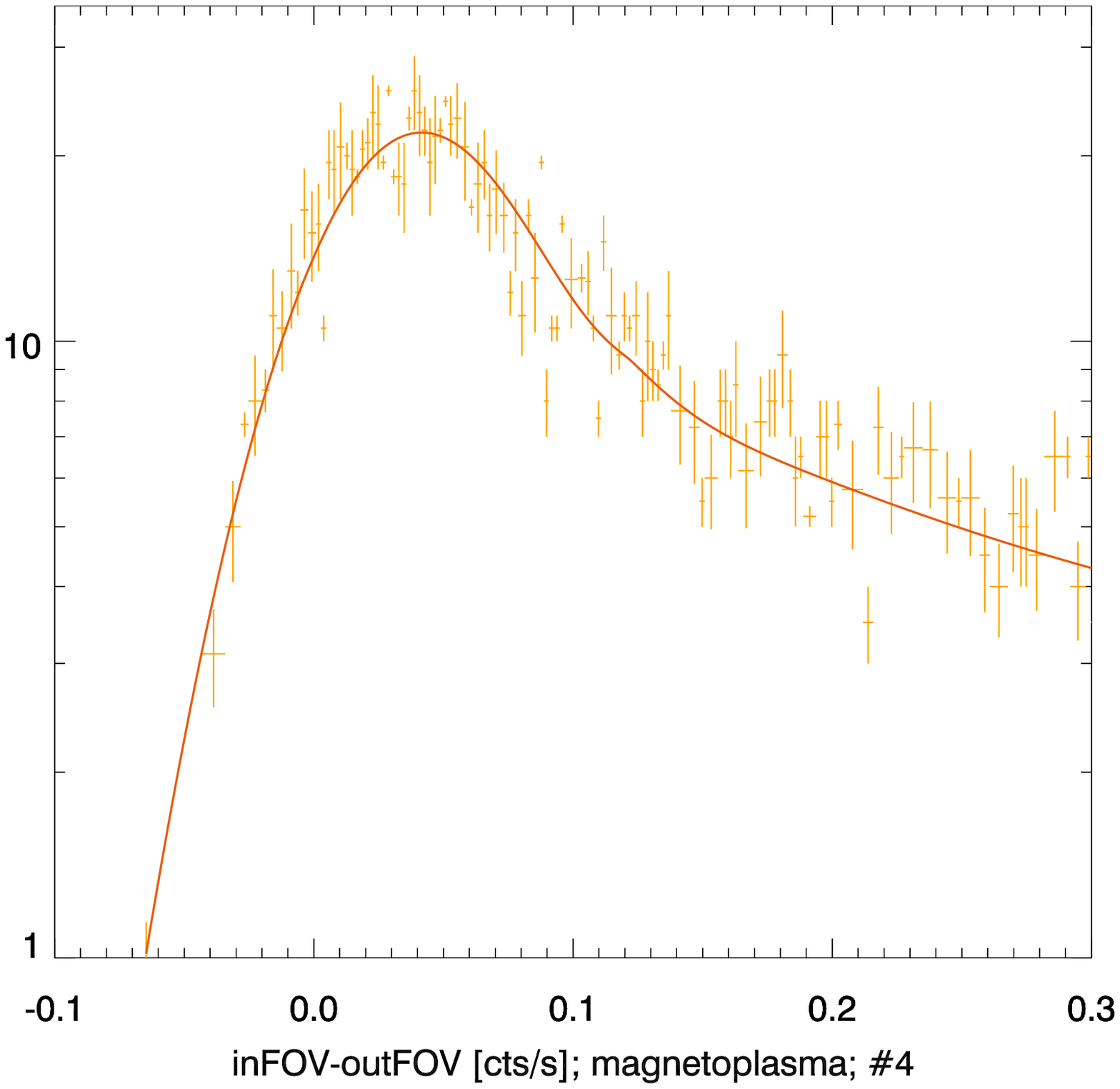}}
 \subfloat{\includegraphics[width=0.5\textwidth]{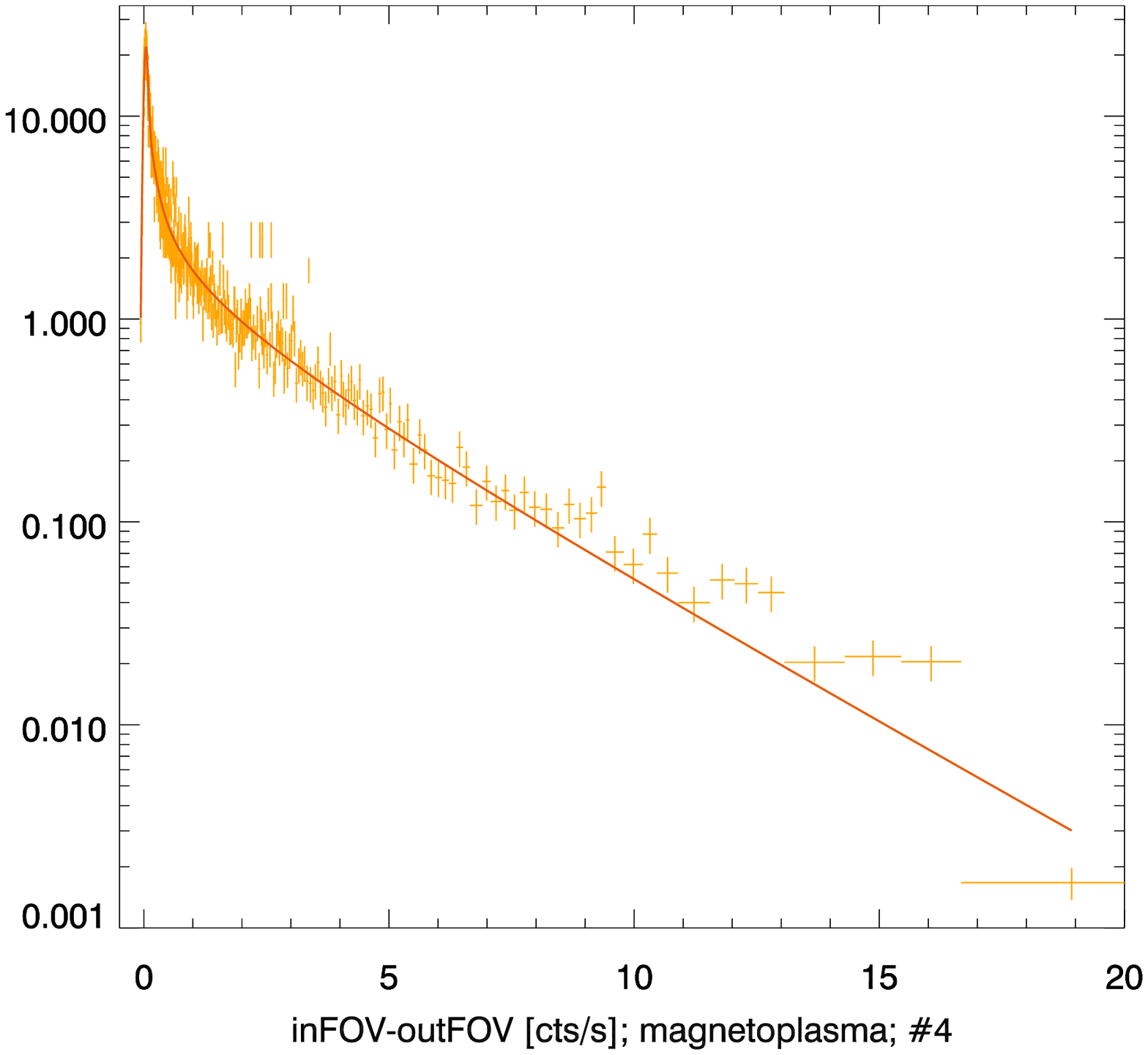} }\\
 \subfloat{\includegraphics[width=0.5\textwidth]{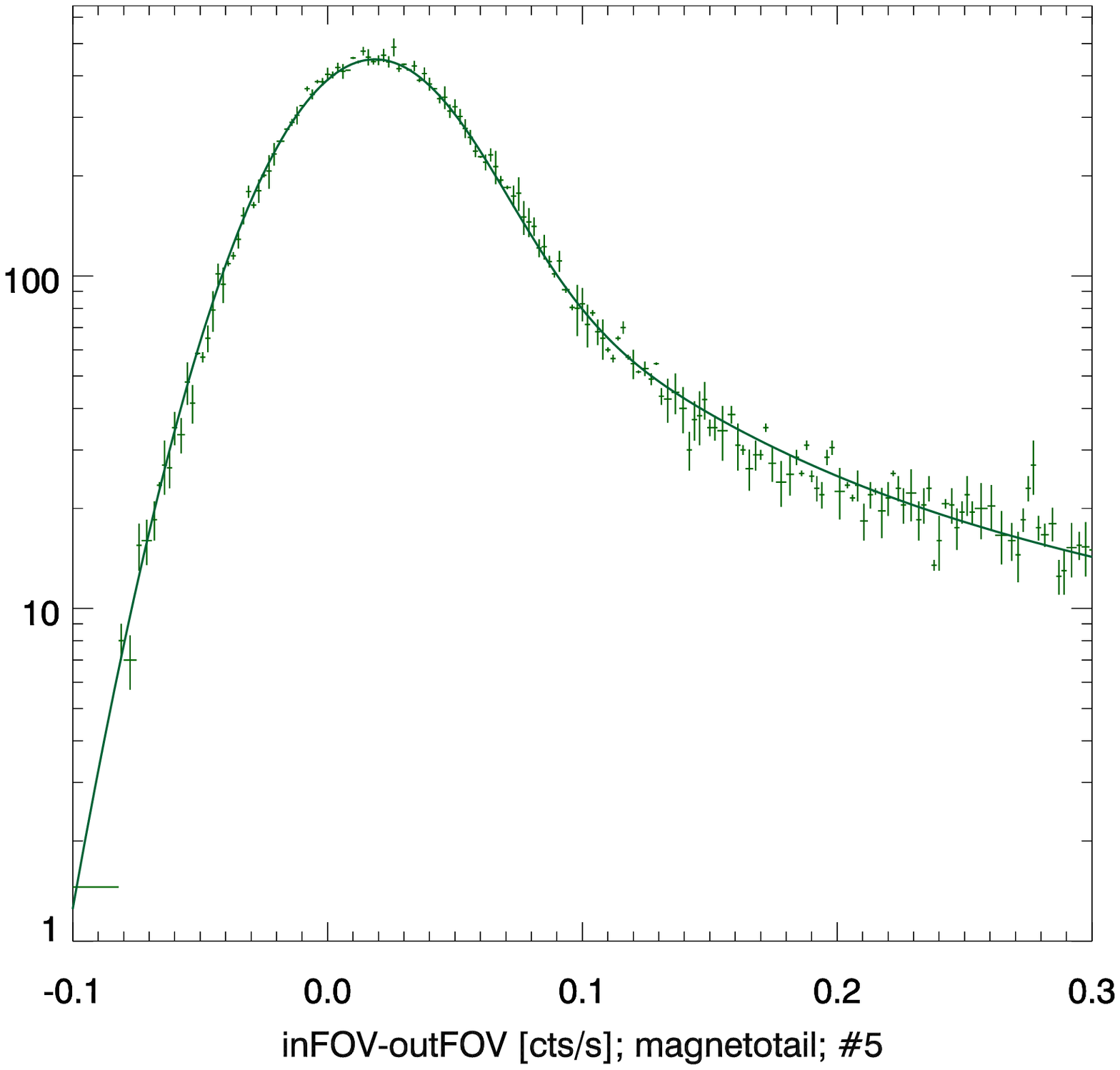} }
 \subfloat{\includegraphics[width=0.5\textwidth]{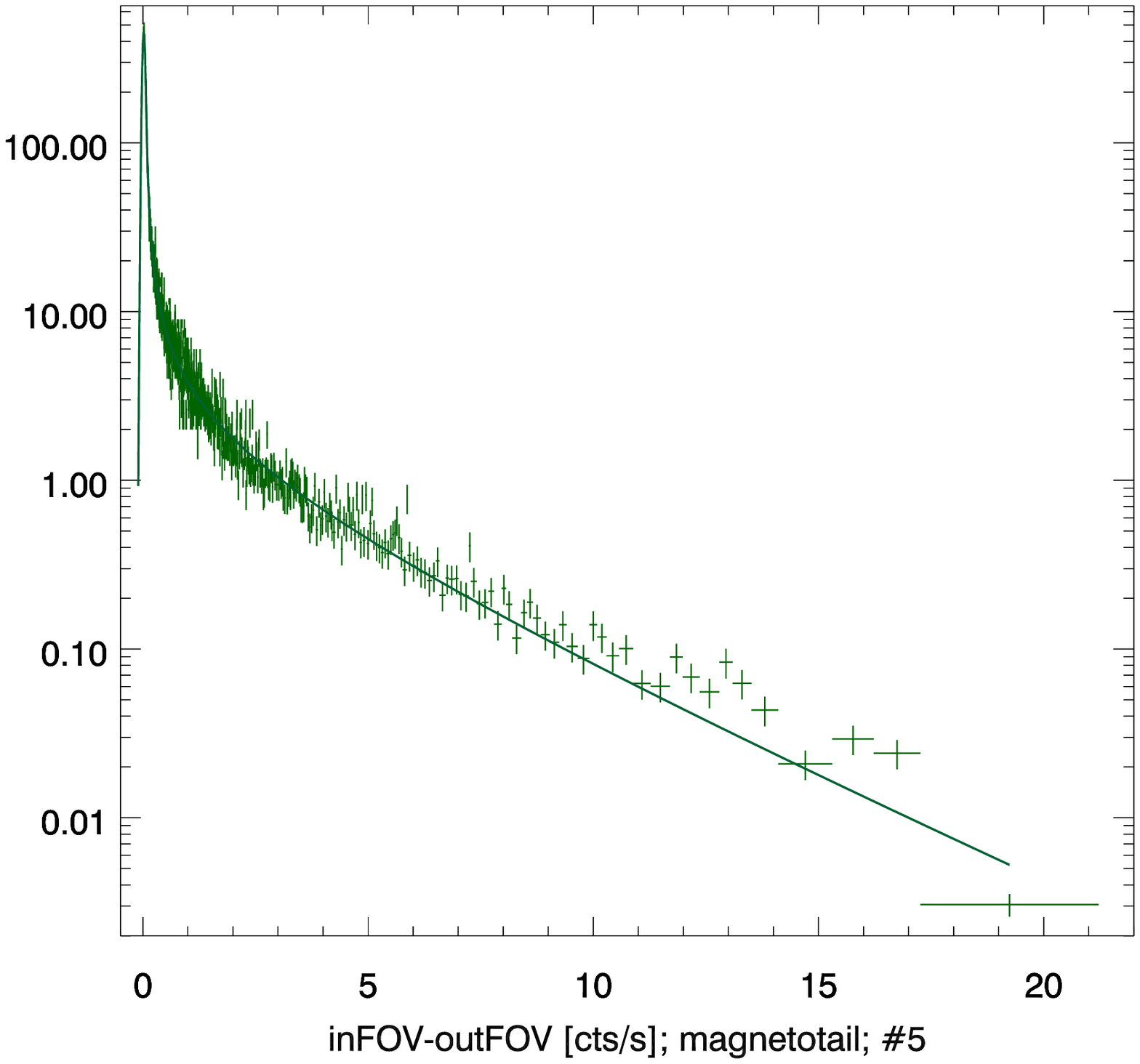} }\\
  \caption{{\it inFOV}-{\it outFOV} distributions of all the magnetozones. Distributions in the left column are zoomed in the range [-0.1,0.3] for a better visualization 
  of the low-intensity component. Best fit functions (see text) are overplotted. Different regions are color coded as in Figure \ref{fig:zones}. (Continues in next page)}
\label{fig:histo_fit}
\end{figure}

\begin{figure}
\ContinuedFloat
 \subfloat{\includegraphics[width=0.5\textwidth]{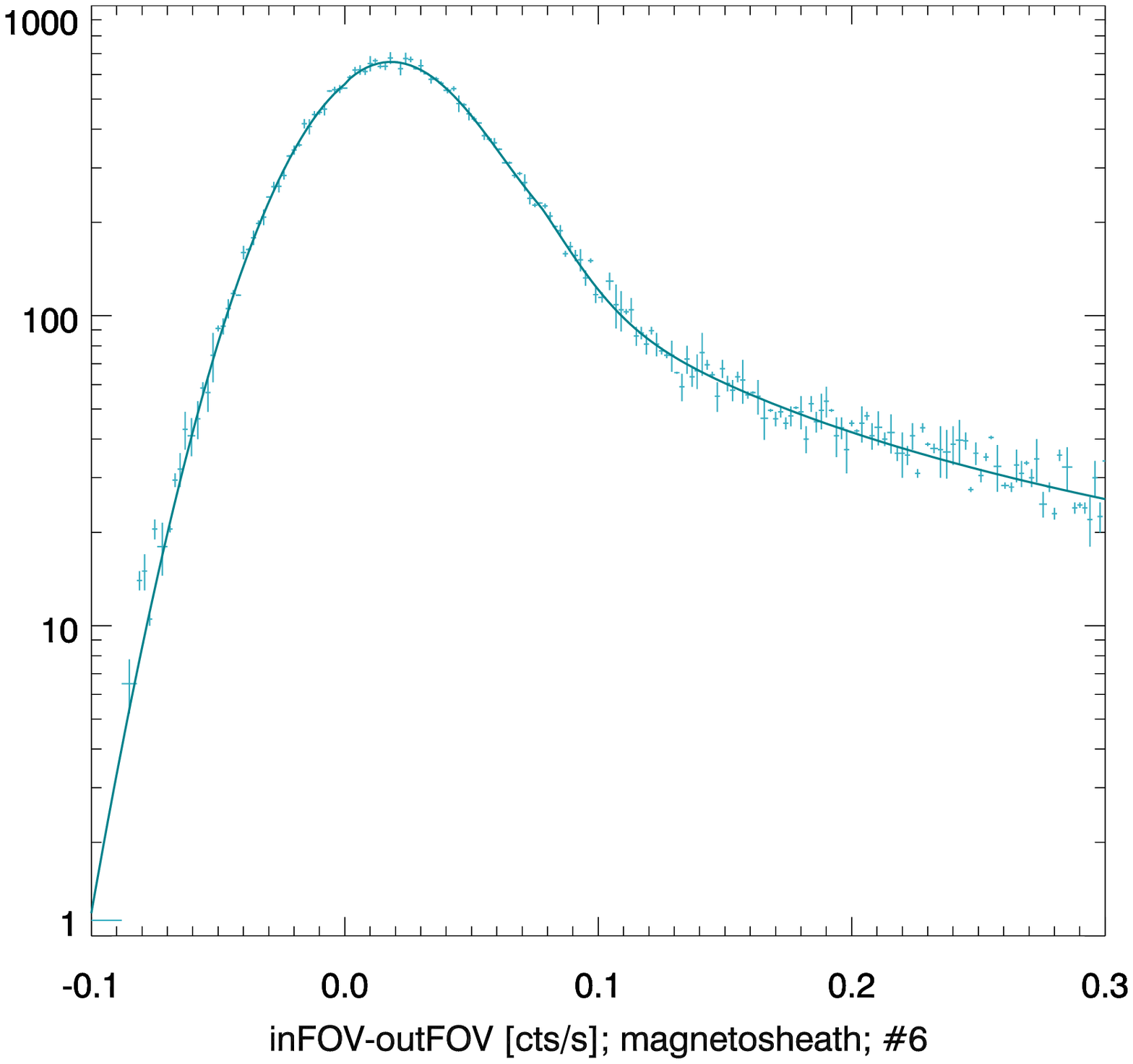}}
 \subfloat{\includegraphics[width=0.5\textwidth]{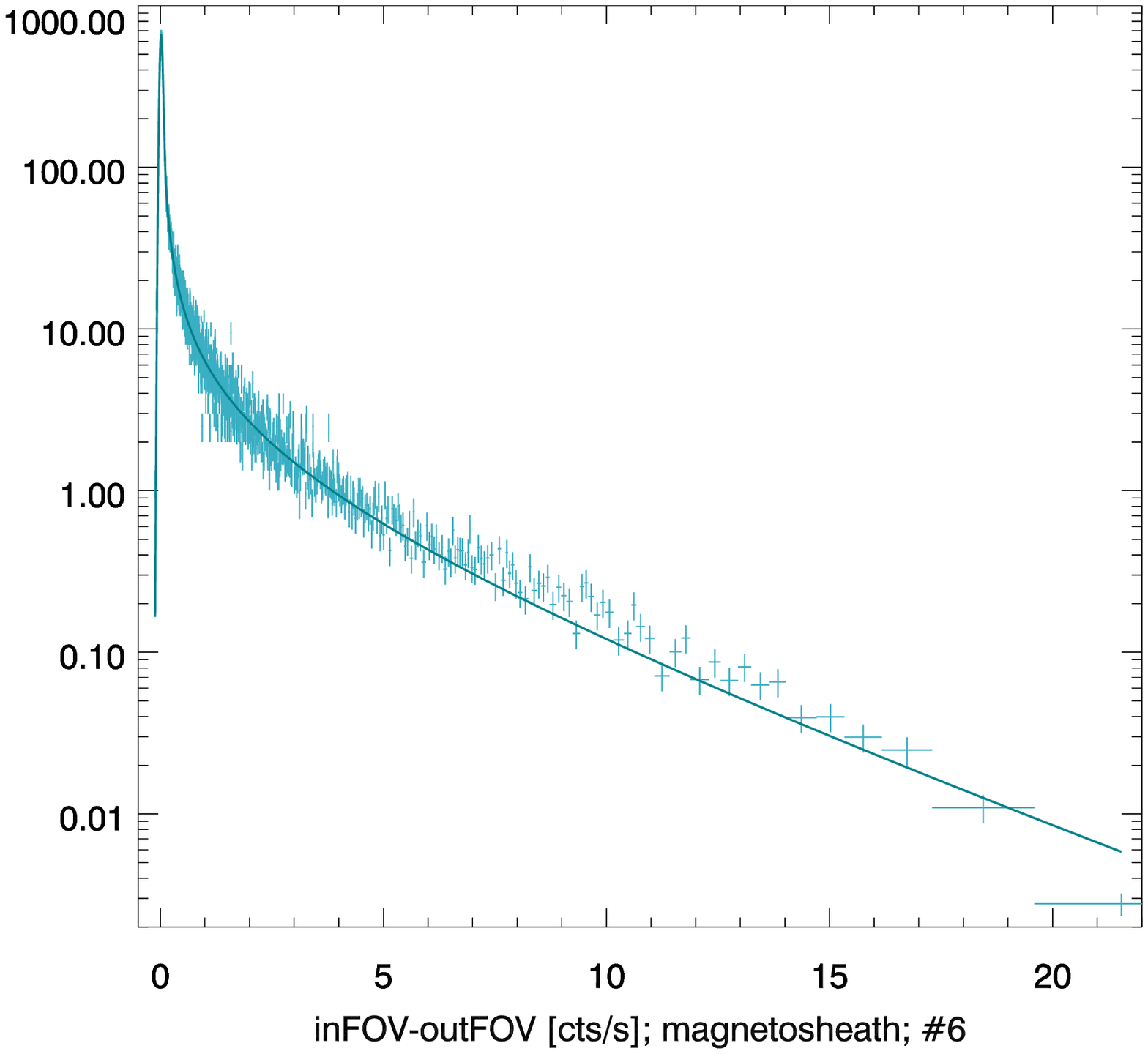}} \\
 \subfloat{\includegraphics[width=0.5\textwidth]{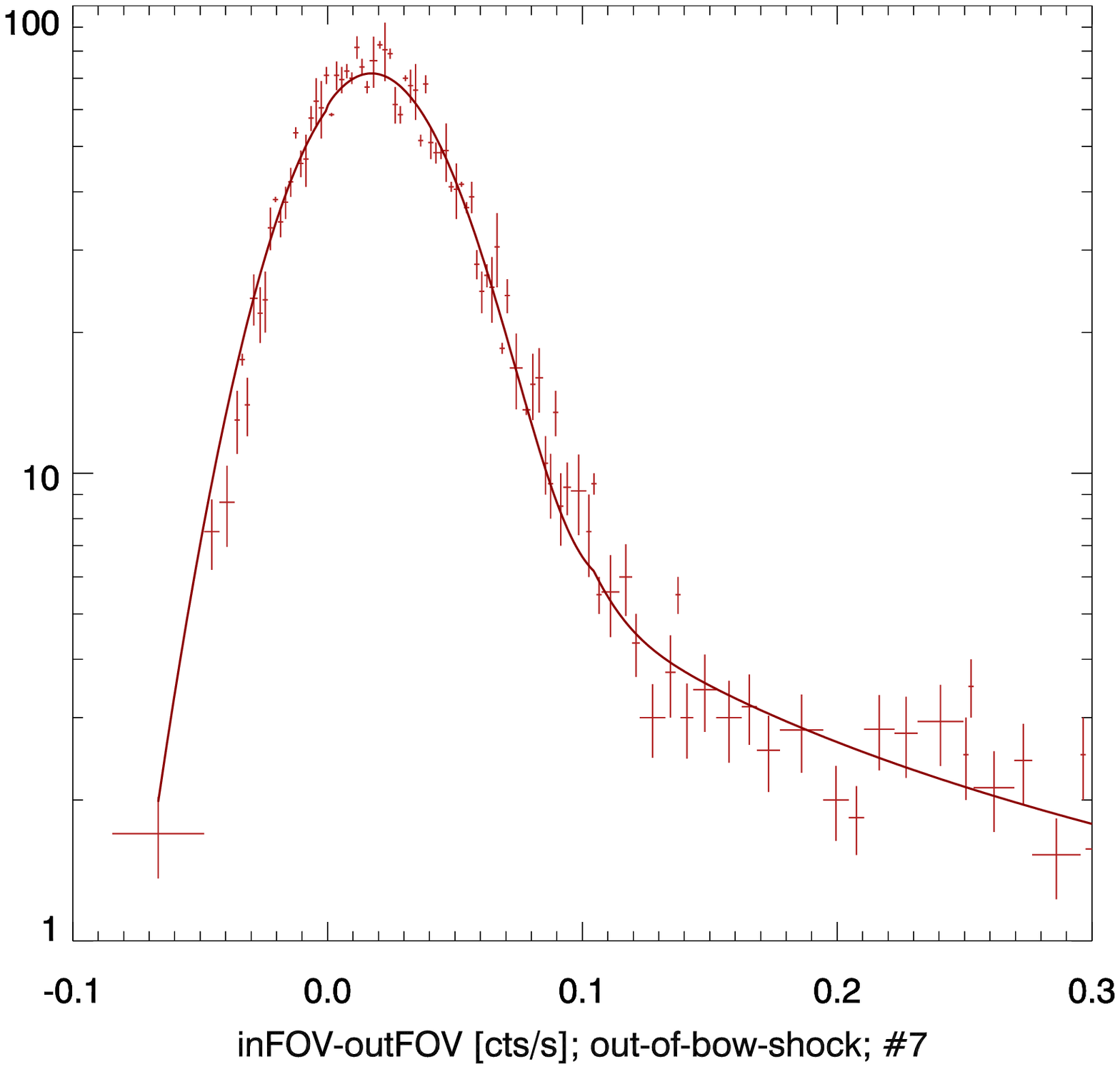}}
 \subfloat{\includegraphics[width=0.5\textwidth]{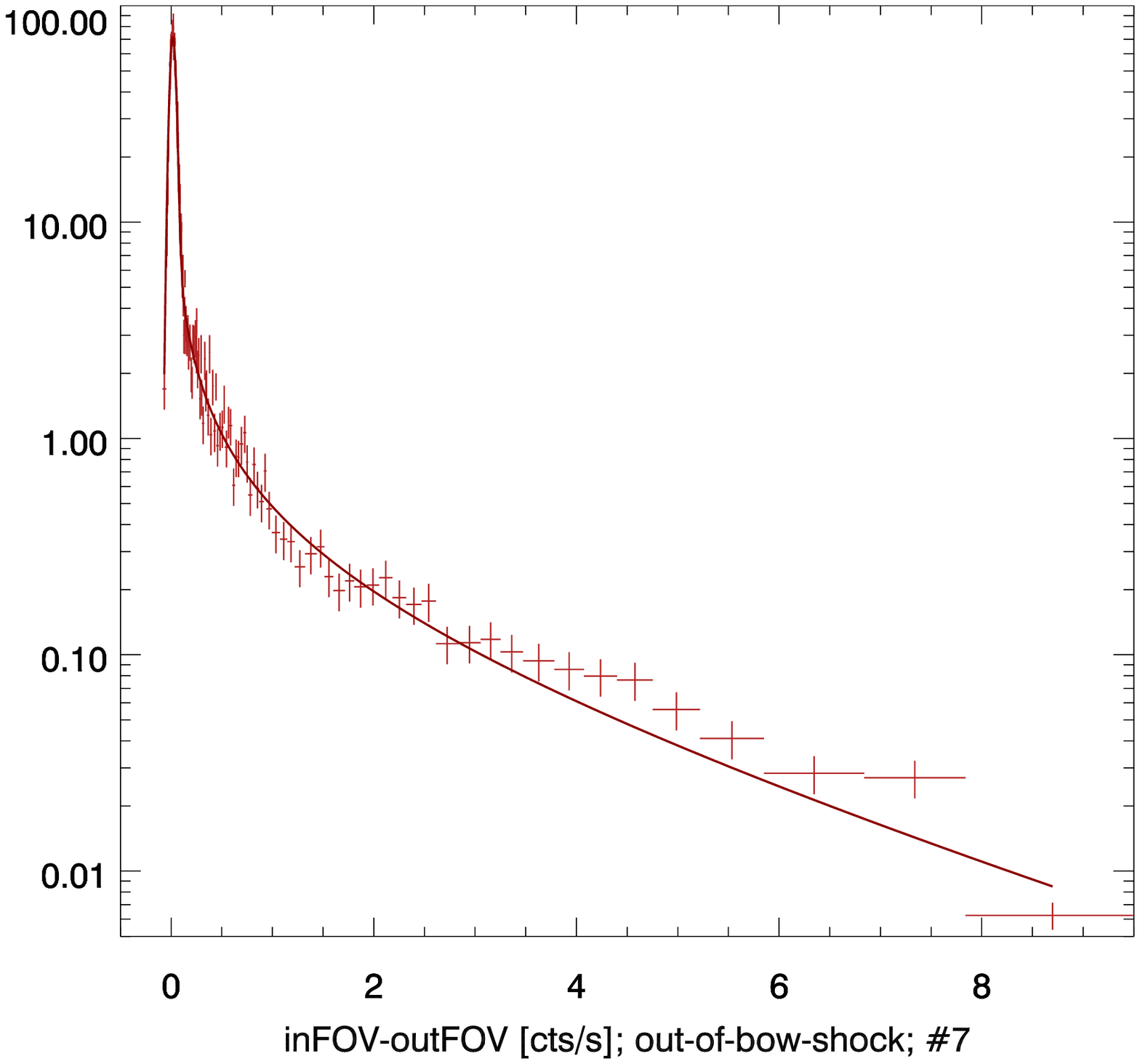}} \\
\caption{(...continued) {\it inFOV}-{\it outFOV} distributions of all the magnetozones. Distributions in the left column are zoomed in the range [-0.1,0.3] for 
a better visualization of the low-intensity component. Best fit functions (see text) are overplotted. Different regions are color coded as in Figure \ref{fig:zones}. }

\end{figure}

\begin{table}
\caption{Gaussian peak positions obtained fitting {\it inFOV}-{\it outFOV} rate; flaring mean rate ({\it inFOV}-{\it outFOV} $>$ 0.1) for each magnetozone.}
\label{table:bestfit}       % Give a unique label
% For LaTeX tables use
\centering
\begin{tabular}{l|c|c}
\hline\noalign{\smallskip}
Magneto-zone & Gaussian Peak &  Flaring mean rate ({\it inFOV}-{\it outFOV} $>$ 0.1) \\
& cts/s & cts/s \\
\noalign{\smallskip}\hline\noalign{\smallskip}
\#3 Plasma sheet & 0.014$\pm$0.003 &  4.075$\pm$0.233 \\
\#4 Magnetoplasma &  0.039$\pm$0.001 & 2.425$\pm$0.037  \\
\#5 Magnetotail & 0.0179$\pm$0.0001 & 1.700$\pm$0.020  \\
\#6 Magnetosheath & 0.0165$\pm$0.0001 & 1.544$\pm$0.015 \\
\#7 Out-of-bow-shock & 0.0168$\pm$0.0002 & 1.522$\pm$0.048 \\
\noalign{\smallskip}\hline
\end{tabular}
\end{table}
We note that peak positions show very modest variations from a zone to another as far as the 
three external zones (\#5, \#6, and \#7 which contain most of the data) are concerned: for 
these regions the magnetic environment has a modest influence on the {\it inFOV} background excess.
The statistics in region \#3 (plasmasheet) is very low and the obtained curve hosts some artificial
features that the fitting procedure introduces to follow distribution irregularities. Best-fit values for this
region, albeit with small error bars, are not reliable from a physical point of view and we cannot use
them to draw any conclusion.
The region \#4 features a higher best-fit value for the peak. 
However in this region the contribution of the flaring component is higher
and it becomes comparable to the low-intensity component, inducing a possible bias on the 
inference of the best-fit value for the Gaussian peak.
It is impossible to disentangle the contamination of the tail on the peak position from a possible real shift of
the low-intensity component; thus the behavior of the low-intensity component in this region is not easily
interpreted.

While the best fit of the gaussian peak is a suitable parameter to describe 
the low-intensity component, 
the best-fit parameters of the Lorentzian function
are not good indicators to quantify the intensity of the flaring
component. A suitable indicator is provided by 
the mean of the high-rate-component: we choose as fiducial threshold 0.1 cts/s and we calculate the
mean value of the {\it inFOV}-{\it outFOV} rate above this threshold and we refer to it as the flaring mean rate. 
The values of the flaring mean rate for each magnetozone are
reported in Table \ref{table:bestfit} (third column).
The flaring mean rates show moderate variations when regions \#5, \#6 and \#7 are concerned, with the out-of-bow-shock 
region featuring the lowest value, though very similar to the other two values. Regions \#3 and \#4 feature higher values. 
% Note that the low statistics in the plasma sheet (\#3)  is significantly high, hence reults concerning this region should be 
% taken with some caution.

%\subsection{Soft protons rate and XMM-Newton altitude}
\subsection{Soft protons flares and XMM-Newton altitude}
\label{sub:altit}

The results reported in the previous section show that the flaring component exhibits modest variations in the various magnetozones,
with magnetoplasma and plasmasheet (which are located in the innermost regions, close to the radiation belts) featuring the highest values, while
the out-of-bow-shock region records the smallest values. This suggests that the {\it inFOV}-{\it outFOV} flux may
be related to the altitude of the satellite rather than to the particular magnetozone.

\begin{figure}
\centering
  \includegraphics[width=0.6\textwidth]{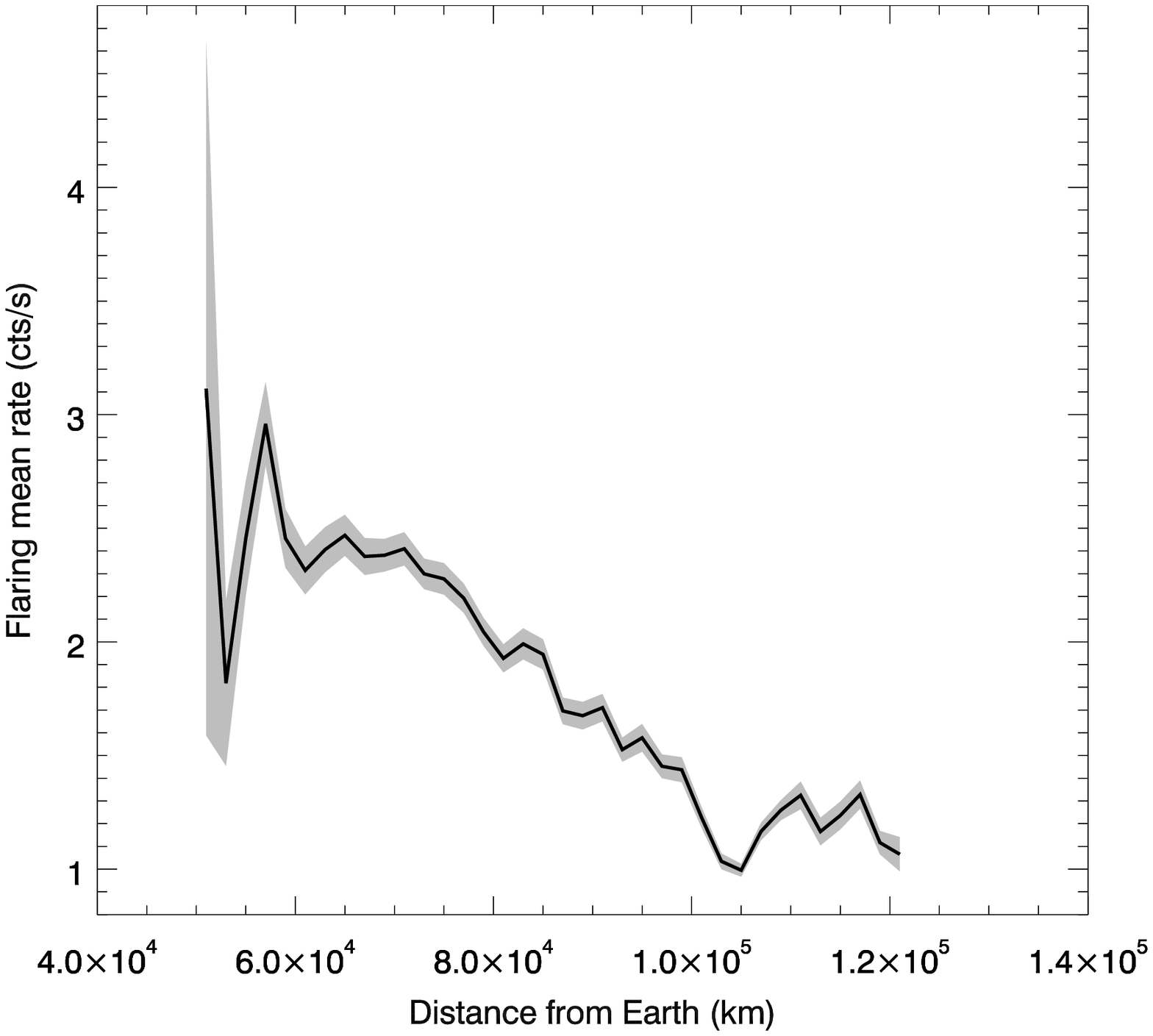}
  \caption{Mean for the {\it inFOV}-{\it outFOV} rate, for count rates $>$ 0.1 cts/s, of the whole sample as a function of the XMM-
Newton distance from the Earth.}
\label{fig:MFR_dist}
\end{figure}

To inspect in detail the {\it inFOV}-{\it outFOV} behavior at different altitudes, we rebinned data using 2-km-wide bins.
The behavior of the low-intensity component cannot be studied through 2-km-wide
shells, since statistics is not enough to perform the fitting procedure. The study of this component requires a specific and extensive analysis 
that is beyond the aim of this article and will be addressed in a forthcoming paper.
For the following discussion we restrict the analysis to the flaring component.
We determined in each bin the mean of the {\it inFOV}-{\it outFOV} rate (for count rates $>$ 0.1
cts/sec, i.e. the flaring mean rate), irrespective of the magnetospheric environment. In Figure \ref{fig:MFR_dist} we plot
this indicator as a function of the XMM-Newton distance from the Earth. The flaring mean rate significantly decreases with the
distance: this means that soft proton flares affect the XMM-Newton background at low altitudes more than at high
altitudes, even though the flaring mean rate never drops below 1
cts/s, showing that this background component can occur in all parts of the XMM-Newton orbit.

\subsection{Soft protons rate Sunward and anti-Sunward}
\label{sub:f-b}

\begin{figure*}
\centering
 \includegraphics[width=0.5\textwidth]{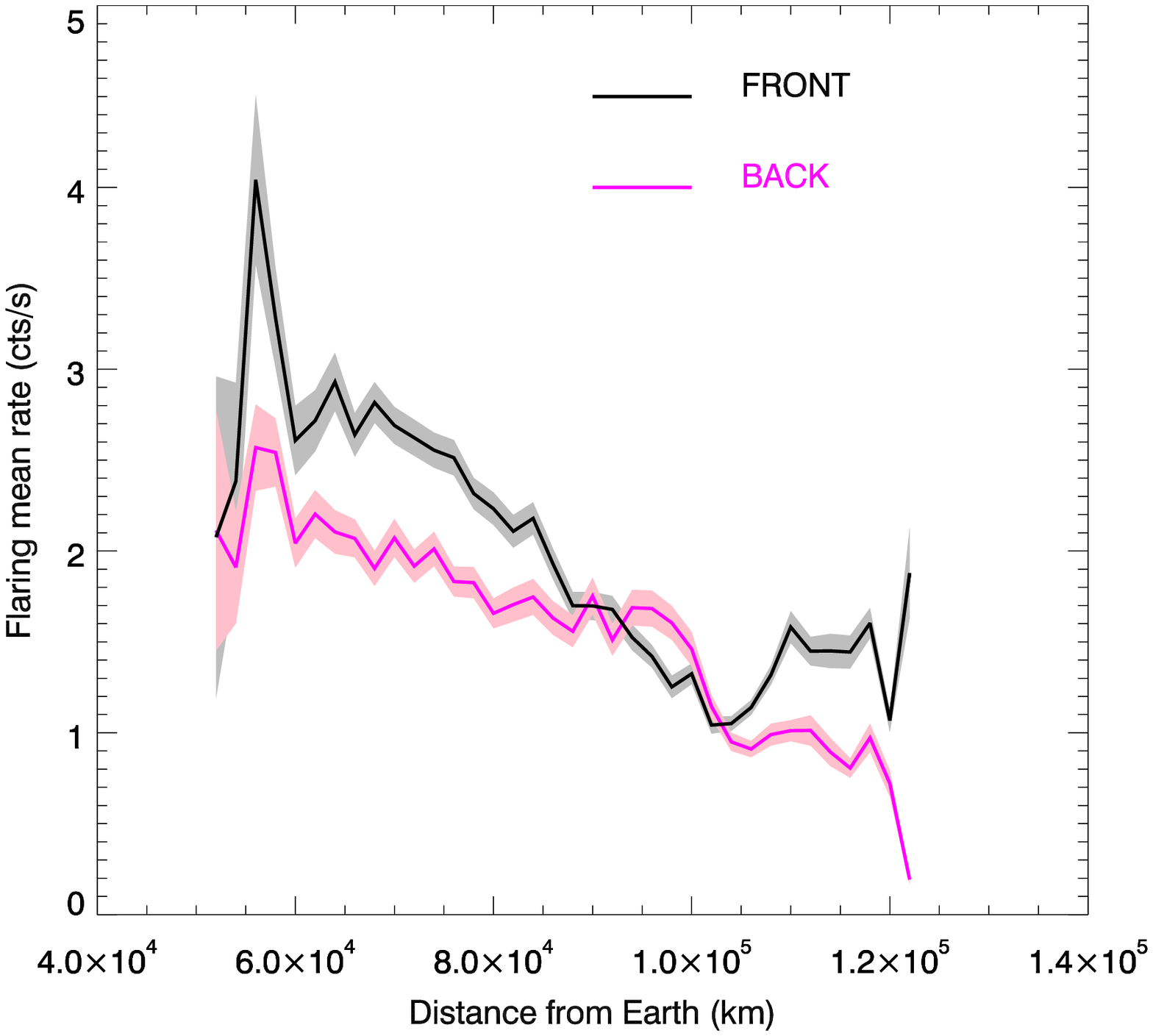}

 \caption{Mean for the {\it inFOV}-{\it outFOV} rate (for count rates $>$ 0.1 cts/s) as a function of distance from Earth in the
dayside (black) and in the nightside (magenta) of the magnetosphere.}
\label{fig:MFR_fb_dist}
\end{figure*}

A further important check concerns the possible differences in the {\it inFOV}-{\it outFOV} rate due to the position of
XMM-Newton in the dayside or in the nightside of the magnetosphere. As mentioned in the previous section, the procedure cannot be applied to 
the low intensity component and we focus on the 
soft proton flares component.
In order to understand if the front/back position with respect to the Sun can be a
discriminatory factor, we evaluate the mean of the {\it inFOV}-{\it outFOV}
rate (for count rates $>$ 0.1 cts/sec) in the same 2-km-wide shells used in Figure \ref{fig:MFR_dist}, and separate
regions Sunward and anti-Sunward. The two profiles are plotted in Figure \ref{fig:MFR_fb_dist}. Both in the dayside and
nightside of the magnetosphere the flaring component features a decrease with the altitude.
In general,
data taken in the dayside have a higher value than data taken in the nightside. This suggests that
regions in the backside of the magnetosphere are less contaminated by soft-proton-flares than regions
in the dayside, with little influence from the magnetospheric environment.

\section{Conclusions}
\label{conclusions}

In this work we studied the role played by the different magnetospheric ambients on the {\it inFOV}
excess background ({\it inFOV}-{\it outFOV} rate) detected by XMM-Newton.
Two main components contribute to the background: a low-intensity component (with 
rate $\lesssim$ 0.1 cts/s) and a soft proton flaring component (with rate $\gtrsim 0.1$ cts/s).
Our analysis shows that moving from a magnetozone to another has a moderate influence both on the
low-intensity background and flaring soft proton component. On the contrary, the soft proton rate is
highly related to the satellite altitude with higher rates at low altitudes. A substantial difference in the
soft proton rate is found when comparing Sunward with anti-Sunward regions, the former featuring a
higher background rate than the latter.

\begin{acknowledgements}
The AHEAD project (grant agreement n. 654215) which is part of
the EU-H2020 programm is acknowledged for partial support.
This work is part of the AREMBES WP1 activity funded by ESA through contract No. 4000116655/16/NL/BW.
Results presented here are based, in part, upon work funded through the European Union Seventh Framework
Programme (FP7-SPACE-2013-1), under grant agreement n. 607452, ``Exploring the X-ray Transient and
variable Sky - EXTraS''.
\end{acknowledgements}

% BibTeX users please use one of
%\bibliographystyle{spbasic}      % basic style, author-year citations
%\bibliographystyle{spmpsci}      % mathematics and physical sciences
\bibliographystyle{spphys}       % APS-like style for physics
%\bibliography{biblio.bib}   % name your BibTeX data base

% NOTA: quando spedisco a EXp Astr non compila la bibliografia. non uso bibtex ma inserico a mano la biblio come da ghizzardi.bbl

% Non-BibTeX users please use

% 
\end{document}